\documentstyle[12pt]{article}

\begin{document}
\title{Impact of localization on Dyson's circular
ensemble}
\author{K. A. Muttalib$^1$ and M. E. H. Ismail$^2$\\
$^1$Department of Physics, University of Florida,
Gainesville, FL 32611.\\
$^2$Department of Mathematics, University of South
Florida, Tampa, FL 33620.\\}
\maketitle
\begin{abstract}
A wide variety of complex physical systems described
by unitary matrices have been shown numerically to
satisfy level statistics predicted by Dyson's circular
ensemble. We argue that the impact of localization in
such systems is to provide certain restrictions on the
eigenvalues. We consider a solvable model which takes
into account such restrictions qualitatively and find
that within the model a gap is created in the
spectrum, and there is a transition from the universal
Wigner distribution towards a Poisson distribution
with increasing localization.\\

\end{abstract}
\newpage
A characteristic statistical property of chaotic (as
opposed to integrable) states in quantum
systems is the distribution of their energies. In
particular,  the
nearest-neighbor spacing distribution or the long
range spectral
rigidity of a local set of levels for a wide variety
of systems in the
chaotic regime agree remarkably well with the
universal Wigner
distributions obtained from the Gaussian random matrix
theory \cite{Mehta,Giannoni}. The same is also
true for ergodic quasienergy eigenstates for a variety of
periodically driven
systems \cite{Reichl} described by the Flouquet
matrix, whose eigenvalues lie on a complex unit
circle, and belong to Dyson's ``circular'' ensemble
\cite{Dyson}.
We will reserve the term Wigner ensemble for
eigenvalues on the real line. Both ensembles follow
the same Wigner distributions in the limit of large
number of eigenvalues.
\par
A new problem in this area is the impact of
localization on the
statistical properties of chaotic eigenstates, which
leads to deviations
from the universal Wigner distributions. While
attempts have been made to generalize the Wigner
ensemble to include such deviations at a
phenomenological level by imposing suitable
constraints \cite{Muttalib,Blecken}, it is clear that
such constraints can not affect the circular ensemble
in the same way
because the eigenvalues are already bounded.
Nevertheless, numerical studies involving scattering
matrix for disordered conductors \cite{Jalabert}
as well as Flouquet matrix for periodically driven
systems  \cite{Jose} show similar
deviations in the spectral
properties \cite{Jung}. It is therefore worthwhile
to consider an analytic model that can accomodate such
deviations in the circular ensemble.

\par
In this paper, by considering the scattering matrix
describing a disordered conductor as an example, we
will argue that the qualitative effect of localization
on the ststistical properties of the circular ensemble
is to provide certain restrictions on the
eigenvalues. We will then construct a solvable model
that takes into account these restrictions in a
qualitative way, and show that this leads to a
transition in the spectral properties from the
universal Wigner distribution towards a Poisson
distribution as a function of a single parameter
related to localization.

\par

Let us consider a one-dimensional scattering of plane
waves of energy $E$ from a potential barrier of width
$a$  and height $V_0$. Define $\hbar k_0=\sqrt{2mE}$
and $\hbar k=\sqrt{2m(E-V_0)}$, $m$ is the mass of the
incident particle. The $2$x$2$ scattering matrix $S$
has the simple form
$$
S=e^{-i\psi}\left(\matrix{
\cos\theta & -i\sin\theta e^{-ik_0 a} \cr
- - - -i\sin\theta e^{+ik_0 a} & \cos\theta \cr
}\right)
$$
where $\cos\theta = 2k_0/\sqrt{4k_0^2+k^2\sin^2(ka)
[1-k_0^2/k^2]}$, $\psi=k_0 a+\mu,\;
\cos\mu=\cos\theta\;\cos ka$. The eigenvalues are
$e^{-i\psi \pm i\theta}$. In a very
crude way, we
might think of the case $E>V_0$ to mimick a metal,
with plane wave states in the region $0<x<a$, while
the case $E<V_0$ will mimick a finite length
insulator with exponentially localized states in
the region. It is clear that while in the former case
the quantity $\cos\theta$ can take on all values from
zero to unity as $k_0$ is varied, it becomes
restricted to values less than unity in the latter
case, where $k=ip$ is imaginary and the term
$k^2\sin^2(ka)[1-k_0^2/k^2]$ is
replaced by $p^2\sinh^2(pa)[1+k_0^2/p^2]$. Such a
restriction can be interpreted as a constraint on the
possible maximum of $Tr(S+S^\dagger)$ which is proportional
to $\cos \theta$,
and the restriction increases with increasing
``localization'' of the waves inside the barrier. In
case of a many-channel quasi
one-dimensional conductor, we can think of the various
channels as having different incoming energies, and an
ensemble of conductors corresponding to different
possibilities for the values of $k$. Channels in the
metallic regime will correspond to having all possible
values of $\theta$ and therefore the eigenvalues will be
uniformly distributed
on the complex unit circle without any
restriction. On the other hand if the channels are
localized, the eigenvalues will be distributed in
a way consistent with the restriction on the trace as
mentioned above. This very
crude argument
suggests that at a phenomenological level, the impact
of localization on the eigenvalue distribution of
scattering matrices can be incorporated by imposing
constraints on $Tr(S+S^\dagger)$. This can be
done in a way suggested by Balian \cite{Balian},
namely introducing Lagrange multiplier functions as
constraints in the joint probability distribution of
eigenvalues. In the present work we will choose a
constraint that has
the qualitative features described above, and for
which one can, at least in principle,
solve for all n-point correlation functions of the
eigenvalue distribution. The hope is that the
qualitative effects obtained from such a solvable
model will be independent of the particular choice
of the model. Indeed we will show that the model
predicts a transition from the highly correlated
Wigner towards an uncorrelated Poisson distribution
in a way that is qualitatively similar to the transition
seen numerically for a variety of systems.

\par\indent
For eigenvalues on the complex unit circle, Dyson's
circular ensemble is based on
the basic ansatz of the random matrix theory that
for a physical
system described by an $N$x$N$ matrix $S$ with
eigenvalues $e^{i\theta_n},
n=1,...N,$ the joint probabilty distribution for
the ensemble of all random $S$  matrices consistent
with given symmetries (unitarity, time reversal etc.)
can be
written quite generally in the form \cite{Mehta}
\begin{equation}
P(\theta _1,.....\theta _N)=\prod_{m<n}
|e^{i\theta _m}-e^{i\theta _n}|^{\alpha}
\prod_{m} w(\theta _m).
\end{equation}
Here $\alpha$ is a symmetry parameter and is equal to
1, 2 or 4 for
orthogonal, unitary and symplectic ensembles
respectively. The function $w(\theta )$ is a Lagrange
multiplier function which might
take care
of any system dependent physical constraint
\cite{Balian}, and in general may depend on
various physical
parameters. Note that for unbounded eigenvalues of
the Wigner ensemble,
such a constraint is required to keep the
distribution normalizable. For the circular ensemble
the above distribution is already normalizable for
$w(\theta)=constant$ and there is in general no need for
additional constraint terms. Dyson has shown
explicitly that the two-level correlation function for
the above distribution for $w(\theta)=\frac{1}{2\pi}$
is identical
to that of the Wigner ensemble for unbounded
eigenvalues in the large $N$ limit, and therefore
leads to the same universal Wigner distributions.
However, this distribution is valid only in the weakly
disordered or chaotic regime, and as we argued before,
the impact of localization can be accomodated
phenomenologically by choosing a Lagrange multiplier
function constraining $Tr(S+S^\dagger)$, or equivalently
$\cos\theta$. Because we have no microscopic model at
this point, we will choose the constraint, with the correct
qualitative features, such that the model is exactly solvable.
\par
Our model corresponds to the choice
\begin{equation}
w(\theta)\sim (1-\cos\theta)^{N/\lambda}.
\end{equation}
Clearly  this has the qualitative features mentioned above,
where the parameter $\lambda$ will serve as a
measure of
localization; decreasing $\lambda$ increases the
constraint on $\cos\theta$. We will show that this model is
solvable in the sense that the spectral correlations can be
written down in terms of known functions. It turns out that
a more general model with two independent parameters, which
contains our model (2) as a special limiting case, is also
exactly solvable. Because of its simplicity as well as
possible relation to
other problems in physics, we will start with the more general
model, write down the general solution, and will come back to
our special limiting case when we analyze and interpret the
solution.
\par
The more general two parameter model is defined by
the choice
\begin{equation}
w(\theta)=\frac{1}{2 \pi}\vert
\frac{(q^{1/2}e^{i\theta};q)_{\infty}}
{(a q^{1/2}e^{i\theta};q)_{\infty}} \vert ^2, \;\;\;\;
 0< q < 1,\;\;\; a^2q < 1,
\end{equation}
where we have used the notation
$(x;q)_n=\prod_{k=0}^{n}(1-xq^k)$.
With the choice $a=q^{N/\lambda}, \lambda >> 1,$ and
$q=e^{-1/N}$ in the limit $N \rightarrow \infty$, or
equivalently
$q \rightarrow 1$, we obtain $w(\theta)=
\frac{1}{2 \pi} 2^{N/\lambda} (1-\cos \theta)^{N/\lambda}$
\cite{Askey}, which is  our model defined in (2). We will
first obtain the general solution for the model (3), and
show that only in the above special limit the impact of
localization becomes observable in the spectral correlations.
In particular we will
show that in this limit a gap appears in the density. We will
also show as an
explicit example that in this case the number variance obtained
from the two-level function shows deviations from
the Wigner distribution, towards a Poisson limit. Note that in
the other limit $a=0$ and  $q \rightarrow 0$,
$w(\theta) \rightarrow \frac{1}{2 \pi}$, and the model reduces
to Dyson's circular ensemble.

\par
For simplicity, we will consider only the case where
the symmetry parameter $\alpha=2$, corresponding to
the case without time reversal symmetry. We use the
method of orthogonal polynomials \cite{Mehta} and
write the product term
$\prod _{m<n}|e^{i\theta _m}-e^{i\theta _n}|$ as a
Vandermonde determinant whose elements form a set of
polynomials orthogonal with respect to the measure
$w(\theta)$. For our particular choice of $w(\theta)$ given in
eqn. (3), these are the (normalized) Szeg\"o polynomials
generalized by Askey \cite{Askey}:

$$
\Phi_n(e^{i\theta};q)=q^{n/2}\left[\frac{(q,q)_n
(q,q)_{\infty} (a^2q,q)_{\infty}}{(a^2q,q)_n
(aq,q)_{\infty}(aq,q)_{\infty}}\right]^{1/2}{S^a}_n,
$$
\begin{equation}
{S^a}_n=\sum_{k=0}^{n}\frac
{(aq;q)_k (a,q)_{n-k}(q^{-1/2}e^{i\theta})^k}
{(q;q)_k (q;q)_{n-k}}.
\end{equation}
The polynomials satisfy the orthogonality relation
\begin{equation}
\frac{1}{2 \pi}\int_{0}^{2 \pi} \Phi_m(e^{i\theta};q)
\overline{\Phi_n(e^{i\theta};q)}w(\theta)d\theta=
\delta_{m,n}
\end{equation}
where the overline denotes complex conjugate.
In terms of these polynomials the two-level
correlation function is given by \cite{Mehta}
\begin{equation}
K_N (\theta,\phi)=\sqrt{w(\theta)}\sqrt{w(\phi)}
\sum_{k=0}^{N-1}\overline{\Phi_k(e^{i\theta})}
\Phi_k(e^{i\phi}).
\end{equation}
We now use the unit circle analog of
Christoffel-Darboux identity \cite{Szego}
\begin{equation}
\sum_{k=0}^{N-1}\overline{\Phi_k(z_1)}\Phi_k(z_2)=
\frac{\overline{\Phi_{N}^*(z_1)}\Phi_{N}^*(z_2)-
\overline{\Phi_{N}(z_1)}\Phi_{N}(z_2)}{1-z_2/z_1},
\end{equation}
where $z_1=e^{i\theta}, z_2=e^{i\phi}$ and we have
used the notation $\Phi_n^*(z)=z^n\Phi_n(1/z)$.
We obtain the large $N$ asymptotics of the polynomials by
noting that the ratio
$\frac{(q;q)_N(aq;q)_{N-k}}{(q;q)_{N-k}(aq;q)_N} =
1+ O(1/N)$. Thus for $N \rightarrow \infty$,
\begin{equation}
\Phi_N(z;q)\approx z^N \sum_{k=0}^{n}\frac
{(a;q)_k (q^{1/2}/z)^k}
{(q;q)_k} =
z^N \frac
{(aq^{1/2}/z;q)_{\infty}}{(q^{1/2}/z;q)_{\infty}}
\;\; , \;\;\;\; a^2q < 1,
\end{equation}
where in the last line we have used the q-binomial theorem
\cite{Gasper}.
The two-level kernel in the large $N$
limit can then be written in the general form

\begin{equation}
K_N \approx \frac{
e^{i(N-1)(\theta-\phi)/2}}{2\pi}\left[\frac
{(q^{1/2}z_1,q^{1/2}/z_2,
aq^{1/2}/z_1,aq^{1/2}z_2;q)_{\infty}}{(q^{1/2}/z_1,
q^{1/2}z_2,aq^{1/2}z_1,aq^{1/2}/z_2;q)_{\infty}}
\right]^{1/2}
\frac{\sin[N(\theta-\phi)/2-\Delta]}
{[\sin(\theta-\phi)/2]},
\end{equation}
where the shift $\Delta$ is given by
\begin{equation}
\Delta=Im \left[ ln \frac{(aq^{1/2}z_1,
aq^{1/2}/z_2;q)_{\infty}}
{(q^{1/2}z_1,q^{1/2}/z_2;q)_{\infty}}
\right],
\end{equation}
and we have used the notation
$(x,y,..z;q)_n=(x;q)_n (y;q)_n ..(z;q)_n$.
For fixed $q$, in the limit $\theta \approx \phi$,
this can be simplified and we obtain
\begin{equation}
\Delta \approx 2(\frac{\theta-\phi}{2})Re\left[
e^{i(\theta+\phi)/2}\sum_{k=0}^{\infty}
\frac{q^{k+1/2}}{1-z_1q^{k+1/2}}-a
e^{-i(\theta+\phi)/2}\sum_{k=0}^{\infty}
\frac{q^{k+1/2}}{1-aq^{k+1/2}/z_1}\right].
\end{equation}
Writing
$1/(1-xq^{k+1/2})=\sum_{l=0}^{\infty}
\left(xq^{k+1/2}\right)^l$ and summing over $k$
first, we obtain the following identity:

\begin{equation}
\sum_{k=0}^{\infty}\frac{q^{k+1/2}}{1-xq^{k+1/2}}=
\frac{q^{1/2}}{1-q}\sum_{l=0}^{\infty}\left(xq^{1/2}
\right)^l \frac{1-q}{1-q^{l+1}}.
\end{equation}
The factor $(1-q)/(1-q^{l+1}) \rightarrow 1$ for
$q<<1$, while it is $1/(l+1)$ in the limit
$q \rightarrow 1$. In both limits the sum can be
explicitly evaluated; it turns out that the result
for $q \rightarrow 1$ contains the $q<<1$ limit,
giving a single expression valid for both limits.
The result, in the limit $\theta \rightarrow \phi$,
is
\begin{equation}
\Delta \approx
(\frac{\theta -\phi }{2})
\frac{1}{1-q}ln \left(\frac{1-2a\sqrt q \cos\theta +a^2 q}
{1-2\sqrt q \cos\theta + q}\right).
\end{equation}
Eqns. (9) and (13)
constitute the solution for large $N$ for the general model
defined by (3), in the limit $\theta \approx \phi$.
\par
We first consider the density of levels given by
$\sigma(\theta)=K_N(\theta,\theta)$. Using (6), (9)
and (13), we get
\begin{equation}
\sigma(\theta)\approx \frac{N}{2\pi}
\left[ 1+\frac{1}{(1-q)N}ln \left(\frac{1-2a\sqrt{q}
\cos\theta +a^2q}{1-2\sqrt{q}\cos\theta +q}
\right)\right].
\end{equation}
Note that the density has a
finite $N$ correction to the uniform density $N/2\pi$
of the circular ensemble. It is clear that in the
$N \rightarrow \infty$ limit, the correction might
survive only in the   $q \rightarrow 1$ limit such
that the product $(1-q)N$ is kept finite. This is
precisely the special limit,  namely $q=e^{-1/N}$
and $a=q^{N/\lambda}$  that defines model (2), and
as we argued in the beginning, this is indeed the
limit where we expect the effect of localization to
become observable in the spectral correlations. In the
rest of our discussions we will restrict ourselves to
this limit only.
\par
The expression (14) for the density of levels has one
apparently very disturbing feature. Although it is
properly normalized to $N$, the density actually becomes
negative for sufficiently small values of $\theta$. In
fact the condition for the density to remain positive for
all values of $\theta$ is that the parameter
$\lambda > \lambda_c =2N(\sqrt{e}-1)$. For
$1 << \lambda << \lambda_c$, the density is positive only
for $\theta > \theta_c$ given by
$2\sqrt{e-1} \sin (\theta_c/2)\sim 1/\lambda$. Thus with
decreasing $\lambda$, i.e. increasing localization,
$\theta_c$ increases.
We will now show that the negative density for
$\lambda < \lambda_c$ implies that there exists a gap in
the spectrum for $\theta < \theta_c$.
\par
In order to understand the density for $\lambda < \lambda_c$,
we will briefly use an alternative approach based on large
$N$ ``coulomb gas'' approximation \cite{Dyson2}. If we write
$w(\theta)=e^{-V(\theta)}$, we can interpret the right hand
side of
eq. (1) as $e^{-H}$, where the effective `Hamiltonian'
$H=\alpha \sum_{m \ne n}ln|2\sin \frac{\theta_m-\theta_n}
{2}|-\sum_n V(\theta_n)$ and the eigenvalues are given by the
stationary condition
\begin{equation}
V^\prime (\theta)=\alpha P\int_I d\phi\sigma(\phi)\cot\frac
{\theta-\phi}{2}\;\; ,
\end{equation}
where $\sigma(\phi)$ is the density to be evaluated, $V^\prime$
is the derivative of $V$ with respect to $\theta$,  $P$ denotes
a principal value integral, and the range $I$ of the integral is
determined from the normalization $\int_I d\phi\sigma(\phi)=N$.
Expanding
$\cot(A-B)$ and using the normalization, we get ($\alpha=2$)
\begin{equation}
V^\prime (\theta)=N\cot\frac{\theta}{2}+\csc^2
\frac{\theta}{2}P\int_{\theta_c}^{2\pi-\theta_c} d\phi
\frac{\sigma(\phi)}{\cot\frac{\phi}{2}-\cot\frac
{\theta}{2}} \;\; ,
\end{equation}
where we have allowed for the possibility that the eigenvalues
lie in the region $|\theta|\ge \theta_c, \;\; \theta_c \le \pi$.
For our model, $V(\theta)=-ln(1-\cos\theta)^{N/\lambda}+const.$.
Using $x=[\cot\frac{\theta}{2}]/[\cot\frac{\theta_c}{2}]$ and
$y=[\cot\frac{\phi}{2}]/[\cot\frac{\theta_c}{2}]$, we can
rewrite eq.(16) as
\begin{equation}
- - - -N(1+\frac{1}{\lambda})\frac{bx}{1+bx^2}=P\int_{-1}^{1}
dy\frac{f(y)}{x-y},
\end{equation}
where we have defined $b=\cot^2(\theta_c/2)$, and
$f(y)dy=\sigma(\phi)d\phi$. This integral can be inverted
\cite{Akhiezer} to give
\begin{equation}
f(x)=-\frac{N(1+\lambda)b}{\lambda \pi^2}\sqrt{\frac
{1-x}{1+x}}P\int_{-1}^{1}\sqrt{\frac{1+y}{1-y}}
\frac{y}{1+by^2}\frac{dy}{y-x}.
\end{equation}

The integral can be evaluated explicitly, giving
$\frac{\pi}{\sqrt{1+b}}\frac{1+x}{1+bx^2}$. Going back to the
original variables, we obtain
\begin{equation}
\sigma(\theta)=\frac{N}{2\pi}\frac{1+\lambda}{\lambda}
\sin\frac{\theta_c}{2}\sqrt{\cot^2\frac{\theta_c}{2}-
\cot^2\frac{\theta}{2}}, \;\;\; |\theta| > \theta_c.
\end{equation}
The normalization condition gives $\sin \frac{\theta_c}{2}=
\frac{1}{1+\lambda}\sim \frac{1}{\lambda}$ for $\lambda >> 1$.
This agrees with our previous result on the existence of the
gap as well as its dependence on $\lambda$.
A similar model, with
$w(\theta)=e^{\frac{2N}{\lambda}\cos\theta}$ has been solved for
the density in the saddle point approximation in the context of
the large $N$ behavior of $U(N)$ lattice gauge theories in two
space-time dimensions \cite{Gross}. A similar gap was found
(at $\theta =\pi$), which suggests that the result is not peculiar
to the particular model we chose; in particular the results from
our solvable model should be qualitatively valid for models
involving qualitatively similar constraints on $Tr(S+S^\dagger)$.
\par
The advantage of our solvable model is that we can
go beyond the density and evaluate the two-level
kernel from which all n-point correlation functions
can be calculated. However, we can not use (5) and (6) directly
because of the gap in the spectrum.
The existence of the gap suggests that we must allow for this
possibility from the beginning, and
replace eqn. (5) by
\begin{equation}
\frac{C}{2 \pi}\int_{\theta_c}^{2 \pi-\theta_c}
\Phi_m(e^{i\theta};q)
\overline{\Phi_n(e^{i\theta};q)}w(\theta)d\theta=
\delta_{m,n}
\end{equation}
Although this means that the polynomials are no longer given
exactly  by (4), we note that for small $\theta_c$, the density
in the large $N$ limit is  almost uniform everywhere except near
the edges. If we restrict ourselves to this uniform density
regime, far from the edges, then the  only real effect of the
gap is to affect the normalization. We have taken this into
account  simply by renormalizing the polynomials (4) by a
factor $\sqrt{C}$ in (20) above. For small values of
$\theta_c$, equivalent to large $\lambda$, the
normalization constant is
$C \approx \frac{1}{1+\frac{c}{\lambda}}$, where $c$ is a
constant $O(1)$. We will restrict our following discussions
only to the regime $\theta \approx \pi$, where the density
is approximately uniform, and the
kernel $K_N(\theta,\phi)$ becomes translationally
invariant:
\begin{equation}
|K_N(\theta-\phi)|\approx \frac{C}{2\pi}\vert\frac
{\sin[\frac{N(\theta-\phi)}{2}(1+\frac{1}
{\lambda})]}{\sin[\frac{\theta-\phi}{2}]}\vert
\end{equation}
where we have included the normalization constant
$C$ explicitly, and $K(\phi-\theta)$ is the complex
conjugate of $K(\theta-\phi)$. In order to compare with the
random matrix theories, we have to ``unfold'' the spectrum
by going to a new variable where the mean spacing between
nearest levels is unity \cite{Mehta}. This is obtained by
choosing the new variables $(\zeta,\eta)=\frac{NC}{2\pi}
(1+\frac{1}{\lambda})(\theta,\phi)$. In terms of
these variables the two-level kernel becomes simply
\begin{equation}
|K(\zeta-\eta)| \approx C\vert \frac{sin\frac
{\pi (\zeta-\eta)}{C}}{\pi (\zeta-\eta)}\vert.
\end{equation}
Note that this looks identical to the two-level kernel of
the gaussian random matrix theory \cite{Mehta}, if we define
a new set of variables $(\zeta^*,\eta^*)=\frac{1}{C}
(\zeta,\eta)$. However, in this new variable the
mean spacing is not unity, but $1/C$, so the
`unfolding' of the spectrum will take us back to the
variable  $(\zeta,\eta)$.
\par
The two-level kernel can now be used to calculate e.g. the
nearest neighbor spacing distribution or the long
range spectral rigidity. To demonstrate the
qualitative effects of localization, we will
explicitly calculate the number variance for an
interval $s$, defined as $(\delta n)^2=< n^2>-< n>^2$.
Using $r=\zeta-\eta$, this is given in terms of the
kernel as \cite{Mehta}
$$
(\delta n)^2=s-2\int_{0}^{s}dr (s-r)|K(r)|^2
$$
\begin{equation}
= s[1-C]+\frac{C^2}{\pi^2}
[ln(2\pi s/C)+\gamma+1]+O(s^{-1}), \;\;\;\;\;\; C=\frac{1}
{1+\frac{c}{\lambda}}\;\;
\end{equation}
where $\gamma$ is Euler's constant. As
$\lambda \rightarrow \infty$, the linear dependence on $s$
cancels exactly and the expression
reduces to the universal logarithmic dependence on
$s$ characteristic of the Wigner distribution.
However, for any finite $\lambda$, there is a leftover
linear dependence on $s$ with the slope increasing
with decreasing $\lambda$ (increasing localization).
This clearly signals a crossover from  Wigner
towards a Poisson (for which $(\delta n)^2=s$)
distribution similar to that seen in the case of
unbounded eigenvalues \cite{Blecken}, and also similar to the
crossover seen in numerical studies of the
number variance for
S-matrix eigenvalues \cite{Jung} describing transport in
mesoscopic conductors \cite{Jalabert} as
well as for $\Delta_3$ statistics
(a related measure of the long range spectral
rigidity \cite{Mehta}) of the Flouquet matrix
eigenvalues describing time evolution of the
Fermi-accelerator model \cite{Jose}. Note that if
$\lambda$ is related to a physical parameter like the
conductance which itself scales with $N$, then starting
from an intermediate case for finite $N$ as given in (23),
the distribution will scale towards either Wigner or Poisson
limit depending on whether
$\lambda$ scales towards  $\infty$ or $0$ with increasing $N$.
\par
We briefly point out that the general model (3) might include
other physically interesting models. For example in the limit
$a=0$ and  $q \rightarrow 1^-$, the function
$w(\theta) \rightarrow \exp[-\frac{1}{1-q}\cos^2\frac{\theta}{2}]$
\cite{Ismail}, which is the model considered in ref. \cite{Gross}.
\par
In summary, we constructed a one-parameter solvable
model (as a special limit of a more general two-parameter solvable
model) for the joint probability distribution of
eigenvalues of unitary matrices which in the large $N$ limit leads
to a gap in the density. The gap increases as a function of the
parameter. By analyzing the effect of the gap on the number
variance, we argued that the model qualitatively describes the
effect of localization.
\par
KAM thanks Y.Chen for valuable comments on the manuscript, and
Z.Qiu for discussion on ref. \cite{Gross}. Research at USF was
partially supported by NSF under grant DMS 9203659.

\par

\end{document}